# Electronic Structure and Superconductivity in $La_{1.55}Sr_{0.45}CuO_4/La_2CuO_4$ Artificial High-$T_c$ Superlattices Probed by Hard and Soft X-ray Spectroscopy

U. M. Jayathilake[1], S. Sheikh[1], T.-L. Lee[2], C. Klewe[3], G. Logvenov[4], G. Campi[5,6,†], A. Bianconi[6,‡], and A. X. Gray[1,*]

[1]*Department of Physics, Temple University, Philadelphia, PA 19122, USA*

[2]*Diamond Light Source Ltd., Didcot, Oxfordshire OX11 0DE, United Kingdom*

[3]*Advanced Light Source, Lawrence Berkeley National Laboratory, Berkeley, CA 94720, USA*

[4]*Max Planck Institute for Solid State Research, 70569 Stuttgart, Germany*

[5]*Institute of Crystallography, Italian National Research Council, IC-CNR, 00015 Rome, Italy*

[6]*Rome International Center for Materials Science Superstripes, RICMASS, 00185 Rome, Italy*

**axgray@temple.edu, †gaetano.campi@cnr.it, ‡antonio.bianconi@ricmass.eu*

## ABSTRACT

In $La_{1.55}Sr_{0.45}CuO_4/La_2CuO_4$ (LSCO/LCO) artificial high-$T_c$ superlattices (AHTS), grown by quantum material design, the superconducting dome can be tuned by varying the geometric ratio $L/d$, where $L$ is the LCO layer thickness and $d$ is the superlattice nanoscale period. Here, we combine hard X-ray photoelectron spectroscopy (HAXPES) and polarization-dependent soft X-ray absorption spectroscopy (XAS) to probe how the electronic structure evolves across the $L/d$-tuned superconducting dome. We observe systematic chemical-potential evolution across the series, together with enhanced spectral weight near the Fermi level, enhanced local and non-local screening signatures, and increased in-plane orbital polarization near the top of the dome. Together, these spectroscopic signatures provide insight into the emergence of Fano-Feshbach resonant superconductivity in the LSCO/LCO AHTS series.

# I. INTRODUCTION

In recent years, breakthroughs in nanoscience and thin-film synthesis have enabled the design of materials with interfacial properties such as superconductivity, magnetism, and two-dimensional electron gases (2DEGs) by precisely engineering their quantum building blocks at the nanoscale [1–5]. A growing body of work has demonstrated that oxide heterostructures can host interfacial superconductivity, establishing the interface as a fertile ground for discovering new quantum states [6–14]. A striking example is superlattices composed of chemically overdoped LSCO and undoped LCO, which host a high-$T_c$ interfacial superconducting state even though neither material is superconducting on its own [8, 15–18]. This emergent state arises from space charge transfer at the LSCO/LCO interface, where holes from the metallic LSCO layer diffuse into the Mott-insulating LCO, creating an ultrathin interfacial region of space charge within stoichiometric LCO. Previous work has demonstrated that this superconducting region is confined to only ~2-3 nm near the interface, extending over a few $CuO_2$ planes [7].

Such interfacial doping sets the stage for quantum confinement and the formation of discrete subbands, which can in turn govern the onset of superconductivity within the Bianconi-Perali-Valletta (BPV) framework [15,16]. BPV theory predicts that when the ratio of the LCO thickness $L$ to the superlattice period $d$ lies in the range $0.3 < L/d < 0.9$, the system is tuned near a Lifshitz electronic transition, where the Fermi surface of an emerging subband hosts an extended van Hove singularity (VHS). In this regime, a Fano-Feshbach resonance is predicted to give rise to resonant multigap superconductivity [15,19,20].

In addition to quantum confinement, the BPV framework predicts that interfacial electric fields further modify the electronic topology of the emerging subbands. Internal electric fields at the LSCO/LCO interface generate Rashba spin-orbit coupling (RSOC), which induces an unconventional Lifshitz electronic topological transition (ETT), where the topology of the emerging Fermi surface evolves from a toroidal (donut-like) shape near the bottom of the subband to a corrugated cylindrical shape [18,21,22]. This transition is predicted to produce a sharp enhancement in the density of states (DOS) near the Fermi level and to strengthen the Fano-Feshbach resonance between pairing channels associated with the toroidal and cylindrical Fermi surfaces, thereby maximizing $T_c$. Within the BPV framework, the Fano-Feshbach resonance between the two superconducting gaps gives rise to a superconducting dome with an asymmetric Fano-like shape centered at the magic ratio $L/d \approx 2/3$ (corresponding to an optimal Sr doping $x \approx 0.15$) [15].

Although transport studies have established the superconducting dome in artificial high-$T_c$ superlattices and the magic ratio $L/d \approx 2/3$, direct spectroscopic evidence for the associated electronic-structure evolution has remained limited. In particular, experimental verification of the chemical-potential ($\mu$) tuning, near-Fermi-level DOS enhancement, interfacial screening mechanisms, and the evolution of occupied and unoccupied electronic states predicted within the BPV framework remains lacking. In this work, using HAXPES [23] and soft XAS, we investigate the electronic-structure evolution underlying the superconducting dome in these AHTS. $\mu$ evolution across the superlattice series is tracked using La $3d$ core-level spectra, while valence-band measurements are used to identify enhancements of near-Fermi-edge ($E_F$) spectral weight. Cu $2p$ core-level spectra are used to probe local and non-local screening channels, providing insight into the evolution of interfacial screening across the superconducting dome. Complementary Cu $L_{2,3}$-edge and O $K$-edge XAS measurements further reveal doping-dependent redistribution of hole states and the evolution of orbital polarization. Our key result is that multiple spectral features evolve systematically as a function of $L/d$, closely tracking the evolution of the superconducting dome, and are consistent with the BPV picture of $\mu$ tuning, near-Fermi-level DOS enhancement, interfacial screening effects, and enhanced orbital polarization.

## II. EXPERIMENT AND MATERIAL CHARACTERIZATION

All four superlattices were grown by molecular beam epitaxy (MBE) on $LaSrAlO_4$ (001) substrates, with the assembly of each monolayer monitored in real time using reflection high-energy electron diffraction (RHEED). Each superlattice consisted of $N = 10$ repeats of an undoped $La_2CuO_4$ layer of thickness $L$ and an overdoped metallic $La_{1.55}Sr_{0.45}CuO_4$ layer of thickness $W$. The structures were designed with a nominal superlattice period $d = L + W = 2.97$ nm, based on growth calibration. The geometric ratio $L/d$ was varied across the series ($L/d$= 0.44, 0.67, 0.78, and 0.89) by adjusting the relative LCO and LSCO thicknesses while maintaining a fixed nominal period, with the LSCO layer adjacent to the substrate and an LCO layer terminating the surface. The average effective hole doping, expressed in terms of the equivalent Sr concentration $x$ in $La_{2-x}Sr_xCuO_4$, is estimated as $x = 0.45(1 - L/d)$, yielding $x_{0.44} = 0.252$, $x_{0.67} = 0.149$, $x_{0.78} = 0.099$, and $x_{0.89} = 0.050$. The effective doping originates from interfacial space charge transfer from the LSCO layers into the stoichiometric $La_2CuO_4$ units.

High-resolution symmetric $\theta$-$2\theta$ X-ray diffraction (XRD) measurements were performed using a Rigaku SmartLab diffractometer equipped with a Ge (220) double-bounce monochromator and Cu $K\alpha_1$ radiation ($\lambda$ = 1.5406 Å). As shown in Fig. 1(a), all four LSCO/LCO superlattices exhibit well-defined out-of-plane 00L diffraction features, including the zeroth-order superlattice reflection ($SL_0$) near $2\theta = 26.8°$ and the first-order superlattice peak ($SL_{-1}$) near $2\theta = 23.1°$. The relatively weak intensity of $SL_{-1}$ is consistent with the small X-ray scattering contrast between the chemically and structurally similar LCO and LSCO layers. A sequence of regularly spaced finite-size fringes is also observed between $SL_{-1}$ and $SL_0$. These fringes arise from coherent interference across the finite number of repeated bilayers. The spectra show the expected number (N − 2 = 8) of thickness fringes for N = 10 superlattice periods, with most fringes clearly resolved despite the experimental noise. The visibility of these fringes and superlattice reflections demonstrates long-range out-of-plane structural coherence and well-defined bilayer periodicity, confirming the high crystalline quality of all four samples.

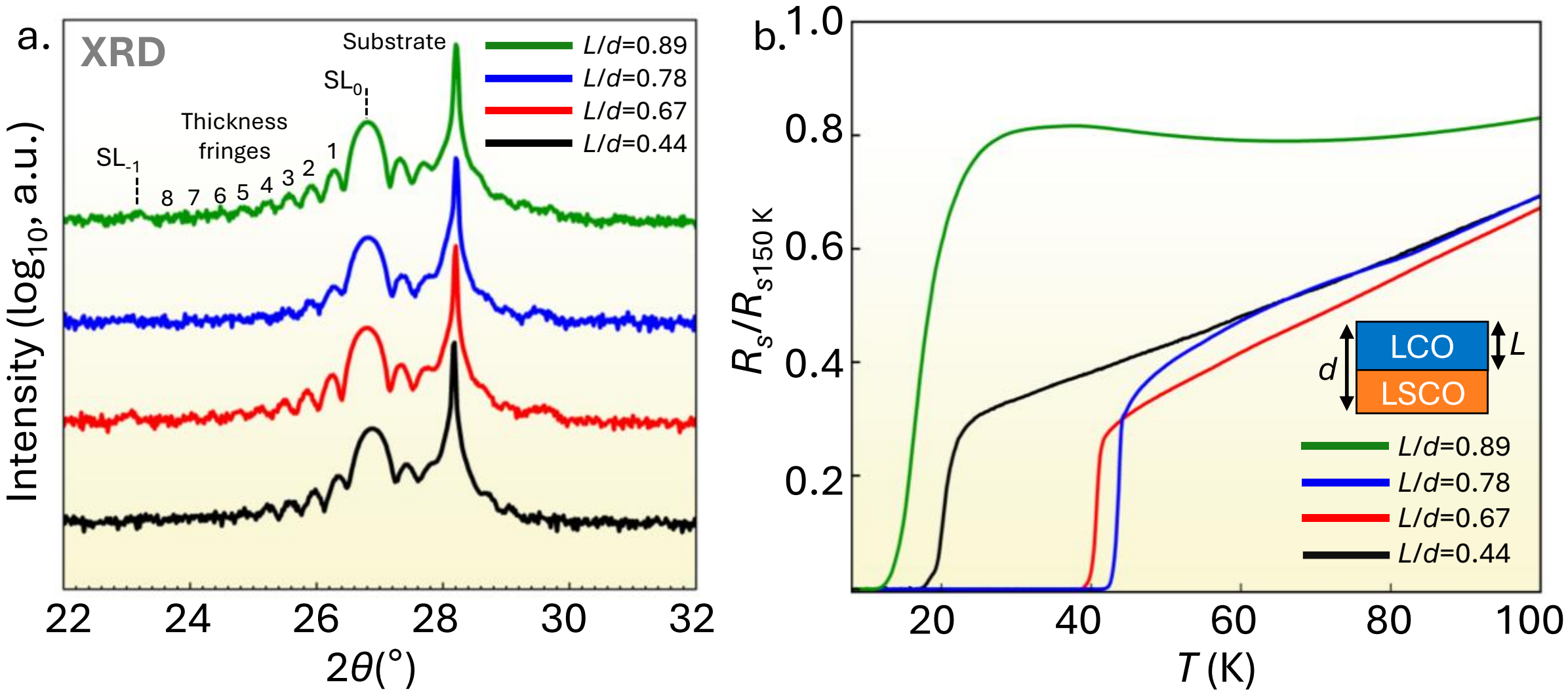

**FIG. 1.** **(a)** High-resolution θ-2θ XRD patterns of the LSCO/LCO superlattice series, showing the zeroth- and first-order superlattice reflections ($SL_0$ and $SL_{-1}$) and well-defined finite-size fringes, indicative of long-range out-of-plane structural coherence and high crystalline quality. **(b)** Normalized sheet resistance $R_s/R_s^{150K}$ versus temperature for LSCO/LCO superlattices with a fixed nominal period and $L/d$ = 0.44, 0.67, 0.78, and 0.89.

Electrical characterization was performed using temperature-dependent resistivity measurements to determine the superconducting transition temperature ($T_c$). Measurements were carried out in a four-point van der Pauw configuration using alternating DC currents of ±10 μA, over a temperature range from room temperature down to 4.8 K. The resulting normalized sheet resistance as a function of temperature for the LSCO/LCO $L/d$ series is shown in Fig. 1(b). The extracted $T_c$ values trace the superconducting dome as a function of $L/d$, reaching a maximum of 43 K for the $L/d$ = 0.78 sample at doping x ≈ 0.1, consistent with the asymmetric Fano-shaped superconducting dome predicted by BPV theory, which is centered at the magic ratio $L/d \approx 2/3$ but whose maximum is offset toward lower doping [15,16].

The HAXPES measurements were performed at 29.5 K and at room temperature using a photon energy of 6.45 keV and a total experimental energy resolution of ~0.3 eV at the high-resolution Surface and Interface Structural Analysis beamline I09 of Diamond Light Source [24]. The energy scale was calibrated using the Fermi edge of a gold (Au) reference measured under identical conditions, and no charging effects were observed. The measured valence-band and core-level spectra, including high-statistics measurements of the Cu $2p$ and La $3d$ core levels, enable detailed analysis of the occupied electronic structure, chemical-potential evolution, and screening channels.

To complement the occupied-state HAXPES measurements, XAS measurements were carried out on the same superlattice samples at 29.5 K at beamline 4.0.2 of the Advanced Light Source [25]. Spectra were collected in luminescence-yield detection mode to probe bulk-sensitive unoccupied electronic states. Cu $L_{2,3}$-edge and O $K$-edge spectra were measured to investigate the unoccupied electronic structure and orbital polarization. Polarization-dependent measurements were conducted with the X-ray electric field vector ($E$) oriented parallel ($E \parallel c$) and perpendicular ($E \perp c$) to the crystallographic $c$-axis. Prior to comparison, the $E \parallel c$ and $E \perp c$ spectra were normalized using the same post-edge intensity reference. X-ray linear dichroism (XLD) was defined as the difference in spectral intensity between the $E \parallel c$ and $E \perp c$ polarization geometries.

## III. RESULTS AND DISCUSSION

### A. Evolution Across the $L/d$ Series

Fig. 2(a) presents a wide-range core-level HAXPES spectrum of the AHTS $L/d$ = 0.44 sample. The spectrum shows well-defined core-level peaks from all constituent elements, including La 3*d* and 3*p*, Cu 2*p*, Sr 3*s* and 3*d*, and O 1*s*. The inset in the bottom right of Fig. 2(a) compares the Sr 3*d* peaks in the shallow-core region of the HAXPES spectra for samples with $L/d$ = 0.44, 0.67, 0.78, and 0.89. The spectra show a systematic decrease in Sr 3*d* intensity with increasing $L/d$ ratio, consistent with the intended reduction in Sr content and the designed variation of the LSCO layer thickness. These observations confirm the expected compositional trend across the superlattice series. The holes introduced by the Sr-substituted LSCO layers provide the interfacial doping that enables emergent superconductivity within the otherwise insulating LCO layers.

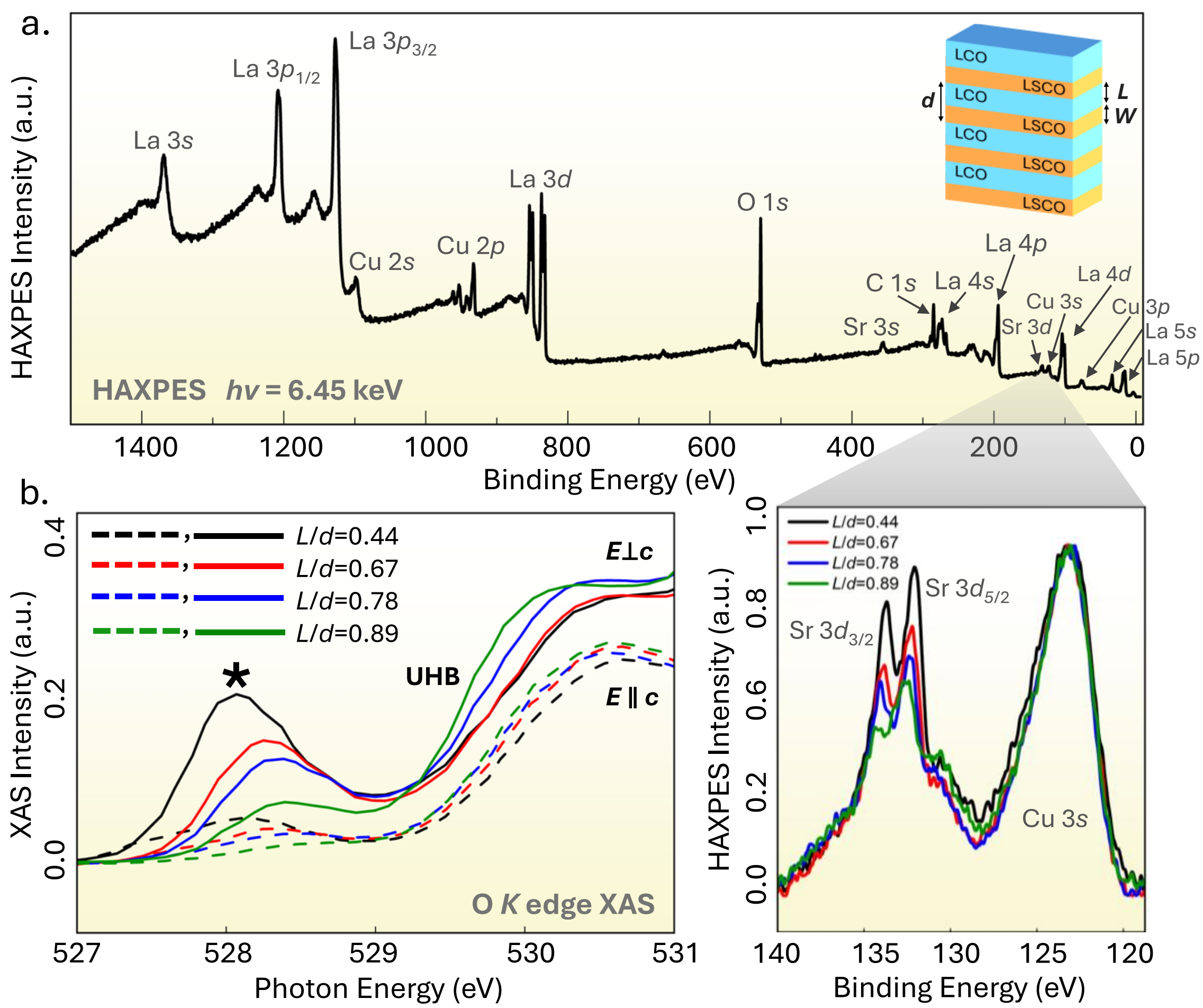

**FIG. 2. (a)** Wide-range HAXPES survey spectrum of the $L/d$ = 0.44 sample, showing core-level peaks from all constituent elements. The structural schematic illustrates the LSCO/LCO superlattice geometry. The bottom-right inset compares the Sr 3*d* shallow-core-level spectra for $L/d$ = 0.44, 0.67, 0.78, and 0.89, demonstrating the systematic decrease in Sr 3*d* spectral intensity with increasing $L/d$. **(b)** O *K*-edge XAS spectra measured in luminescence-yield mode for LSCO/LCO superlattices with $L/d$ = 0.44, 0.67, 0.78, and 0.89. Solid lines correspond to spectra measured with $E \perp c$ polarization, while dashed lines represent spectra measured with $E \parallel c$ polarization.

Fig. 2(b) shows the O $K$-edge XAS spectra measured in a bulk-sensitive luminescence-yield detection mode for LSCO/LCO superlattices with $L/d$ = 0.44, 0.67, 0.78, and 0.89. Solid lines correspond to spectra measured with $E \perp c$ polarization, while dashed lines represent spectra measured with $E \parallel c$ polarization; both polarizations were measured under otherwise identical experimental conditions. The low-energy pre-edge feature, labeled *, is attributed to transitions from the O 1$s$ core level into hole states formed by strong hybridization between O 2$p$ and Cu 3$d$ orbitals and serves as a quantitative measure of doped holes in the $CuO_2$ planes [26]. The systematic variation in the intensity of feature * indicates that, within the AHTS series, the $L/d$= 0.44 and 0.89 samples are effectively overdoped and underdoped, respectively, while the $L/d$ = 0.67 and 0.78 samples lie near the optimal doping regime for AHTS. The spectra further indicate that these holes reside predominantly in the in-plane O 2$p$ orbitals of the $CuO_2$ planes, with only a minor contribution from apical oxygen states.

The higher-energy feature labeled UHB corresponds to transitions into the upper Hubbard band, which dominates the pre-edge region in the undoped charge-transfer-insulating state [27]. Across the superlattice series, the O $K$-edge spectra show a systematic transfer of spectral weight from the UHB to feature * as the $L/d$ ratio decreases, corresponding to an increasing LSCO fraction and higher effective hole doping. This evolution reflects the progressive depletion of Mott-insulating LCO spectral weight and the emergence of doped, metallic $CuO_2$-plane states via this interfacial charge transfer.

Together, the XRD, HAXPES, and soft XAS measurements establish the structural and compositional integrity of the artificial superlattices and provide direct spectroscopic evidence for the evolution of doped-hole states across the series, linking the designed superlattice geometry to the doping-dependent electronic structure.

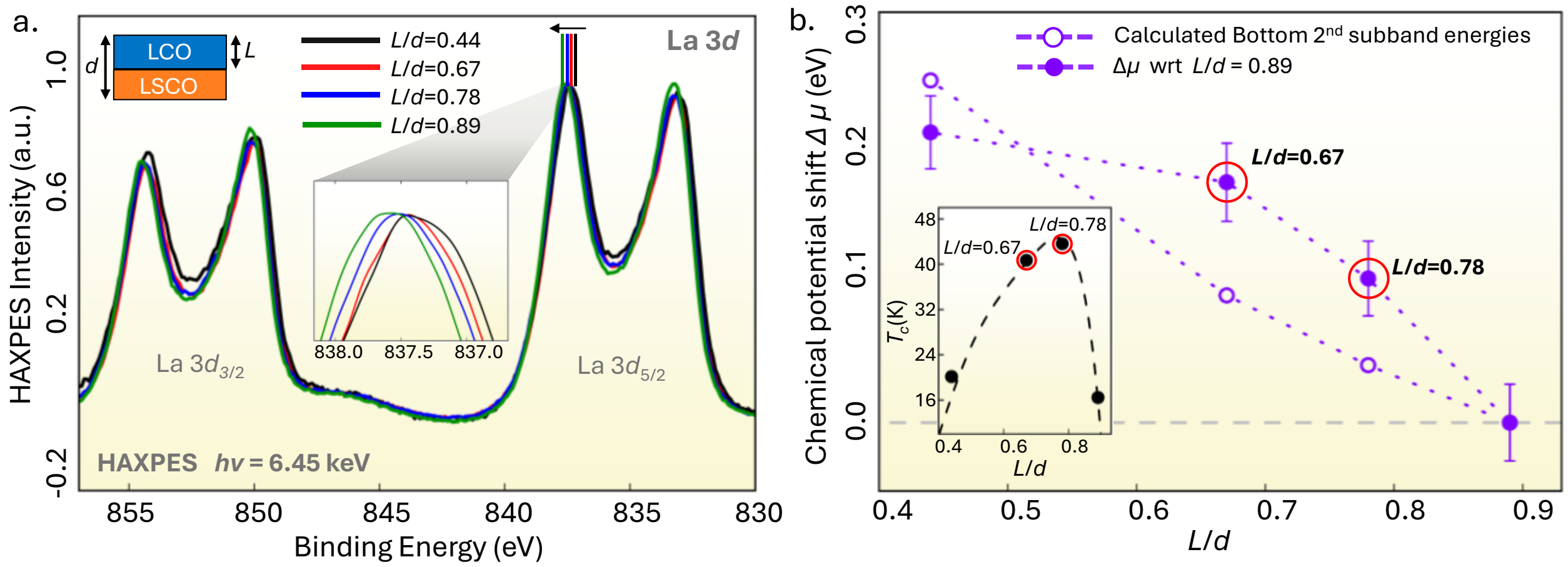

**FIG. 3. (a)** La 3$d$ core-level HAXPES spectra of LSCO/LCO superlattices with different $L/d$ ratios and a fixed nominal superlattice period. The arrow marks the systematic La 3$d$ binding-energy shift across the series, resolved in the zoomed-in inset. **(b)** Chemical-potential shift $\Delta\mu$, extracted from the La 3$d_{5/2}$ core-level peak positions using the $L/d = 0.89$ sample as the reference. Violet (closed) symbols represent the experimentally extracted $\Delta\mu$ values, while violet (open) symbols show the BPV-predicted bottom energies $E_L$ of the second subband for comparison, referenced to the $E_L$ of $L/d$ = 0.89 and reproduced with permission from Bianconi *et al.* [16]. The inset shows the asymmetric superconducting dome of the AHTS series, with symbols corresponding to experimental $T_C$ values and the dashed black line tracing the BPV-calculated rising and falling edges of the dome. The $L/d = 0.78$ and 0.67 samples lie near the maximum of the superconducting dome.

### B. Chemical-Potential Evolution and Second-Subband Occupation

Having established the compositional integrity and doping-dependent electronic structure of the AHTS samples, we next examine $\mu$ evolution across the superlattice series using the La 3$d$ core-level spectra [Fig. 3(a)]. The La 3$d$ binding energy provides a reliable proxy for relative $\mu$ shifts because La atoms reside in the LaO building blocks, where final-state screening effects are significantly weaker than for core levels more directly coupled to the conducting $CuO_2$ planes, such as O 1$s$ and Cu 2$p$ [28,29]. Under the reasonable assumptions that the La valence remains unchanged and that Madelung-potential variations are small over this doping range, the observed La 3$d$ binding-energy shifts provide a direct estimate of $\Delta\mu$ [28].

The extracted $\Delta\mu$ values [Fig. 3(b)] evolve smoothly with increasing hole concentration, showing no evidence of pinning. This behavior is consistent with recent work by Nagamoto *et al.* [28], which suggests gradual filling of the dispersive quasiparticle bands rather than $\mu$ pinning. The continuous evolution of $\mu$ across the AHTS series provides an experimental basis for evaluating whether the system approaches the resonance condition predicted by BPV theory. Within the BPV framework, superconductivity is maximized when $\mu$ approaches the bottom of the second subband, corresponding to an ETT where a second Fermi surface appears. This condition produces a significant enhancement of the DOS due to an extended VHS and activates a Fano-Feshbach resonance between pairing channels, thereby enhancing the superconducting critical temperature at the interface.

As shown in Fig. 3(b), we compare the experimentally determined $\Delta\mu$ values to the theoretically predicted bottom energy ($E_L$) of the second subband for a nominal superlattice period of 2.97 nm within the BPV framework. We find that the $\Delta\mu$ values for the AHTS with $L/d$ = 0.67 and 0.78 lie above $E_L$, indicating that $\mu$ has crossed into the second subband, whereas the $L/d$ = 0.44 sample remains below this threshold. Within the BPV framework, occupation of the second subband is expected to enhance the DOS near the Fermi level, strengthen screening channels, and increase orbital polarization, as discussed in Secs. III C, III D, and III E. Because the AHTS with $L/d$ =0.67 and 0.78 lie near the top of the superconducting dome [see inset of Fig. 3(b)], they are expected to exhibit the strongest enhancements of the near-Fermi-level DOS, screening response, and orbital polarization within the series.

### C. Near-$E_F$ Density-of-States Enhancement

To investigate the DOS enhancement anticipated from the chemical-potential evolution discussed above, we employed bulk-sensitive valence-band HAXPES to track the evolution of the occupied DOS near $E_F = 0$ across the LSCO/LCO superlattice series. As shown in Fig. 4(a), all samples exhibit a near-$E_F$ spectral-weight feature associated with Cu 3$d$-derived states [Figs. 4(c) and 4(e)], consistent with interfacial charge transfer inducing finite occupied spectral weight at $E_F$ across the entire series, including the effectively underdoped members, at both 29.5 K and room temperature. Further analysis of the valence-band maximum (VBM) evolution [Figs. 4(b) and 4(d)] reveals a systematic redistribution of occupied O 2$p$-Cu 3$d$ hybridized states with hole doping. With decreasing $L/d$ ratio, the VBM shifts toward $E_F$, accompanied by an increase in spectral weight, indicating that increasing hole doping drives the antibonding states closer to the

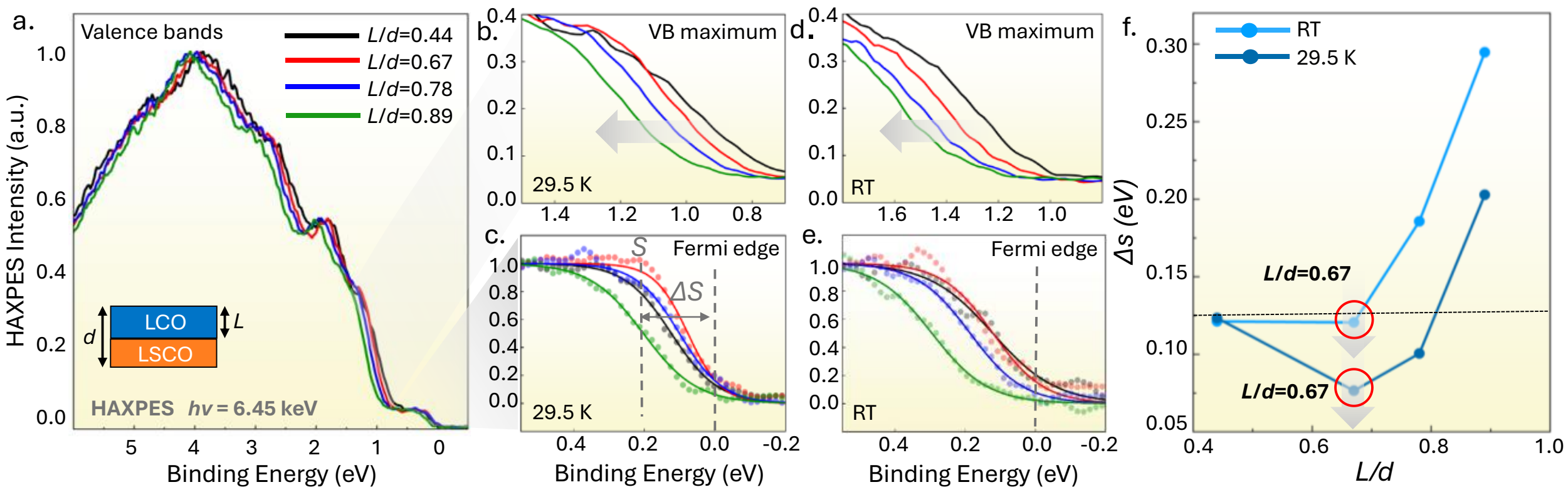

**FIG. 4. (a)** Valence-band HAXPES spectra of LSCO/LCO superlattices with different $L/d$ ratios, normalized to the peak maximum. **(b)**, **(d)** Expanded views of the valence-band maximum (VBM) region measured at 29.5 K and room temperature (RT), respectively. **(c), (e)** Expanded views of the near-$E_F$ valence-band region measured at 29.5 K and RT, respectively, with intensities normalized between 0 and 1. Dots represent the raw HAXPES data and lines show the Fermi-Dirac fits. **(f)** Leading-edge shift $\Delta S = S - E_F$, where $S$ denotes the fitted leading-edge position and $E_F$=0, as a function of $L/d$, extracted by fitting the spectra with Fermi-Dirac distribution functions as in panels (c) and (e) for measurements at RT and 29.5 K.

Fermi level [30]. Within the BPV framework, this upward shift is consistent with the $\mu$ evolution required to approach the Fano-Feshbach resonance condition between pairing channels. The systematic shift of the VBM toward $E_F$ motivates a more quantitative examination of the near-$E_F$ spectral weight.

Although the near-$E_F$ spectral-weight distributions appear comparable across the series, fitting the spectra with Fermi-Dirac functions, as shown in Figs. 4(c) and 4(e), enables a quantitative determination of the leading-edge shift for each AHTS sample, as shown in Fig. 4(f). This analysis reveals that the AHTS with $L/d$ = 0.67 exhibits the smallest $\Delta S$, corresponding to the leading edge closest to $E_F$, indicative of enhanced near-$E_F$ spectral weight at both temperatures. The resulting nonmonotonic evolution indicates that the superlattice geometry tunes the chemical potential, $\mu$, of the $L/d$ = 0.67 sample, which lies near optimal doping, into a regime of enhanced DOS near $E_F = 0$. This behavior is consistent with the enhanced DOS expected as the $\mu$ approaches a VHS-like condition within the BPV framework.

### D. Enhanced Cu 2*p* Screening Response Near the Superconducting Dome Maximum

We examined the HAXPES Cu $2p$ core levels [Fig. 5(a)] to probe screening efficiency across the $L/d$ series. These spectra are highly sensitive to the local and interfacial electronic environment, where the balance between local and non-local screening channels is expected to evolve as the system approaches an ETT. Fig. 5(a) displays the Cu $2p_{3/2}$ and Cu $2p_{1/2}$ core-level components, separated by approximately 20 eV due to spin-orbit coupling. The higher-binding-energy satellite features near ~942 eV and ~962 eV are attributed to Cu $2p^5 3d^9$ final-state atomic multiplet configurations involving $2p$ core holes and $3d$ valence holes. The lower-binding-energy screened features near ~932 eV and ~953 eV have been assigned to Cu $2p^5 3d^{10}\underline{L}$ final-state configurations, where $\underline{L}$ denotes a ligand hole [31,32]. These screened final states arise from configuration interaction between Cu 3$d^9$ and Cu 3$d^{10}\underline{L}$ configurations in the initial state of the undoped Mott insulator.

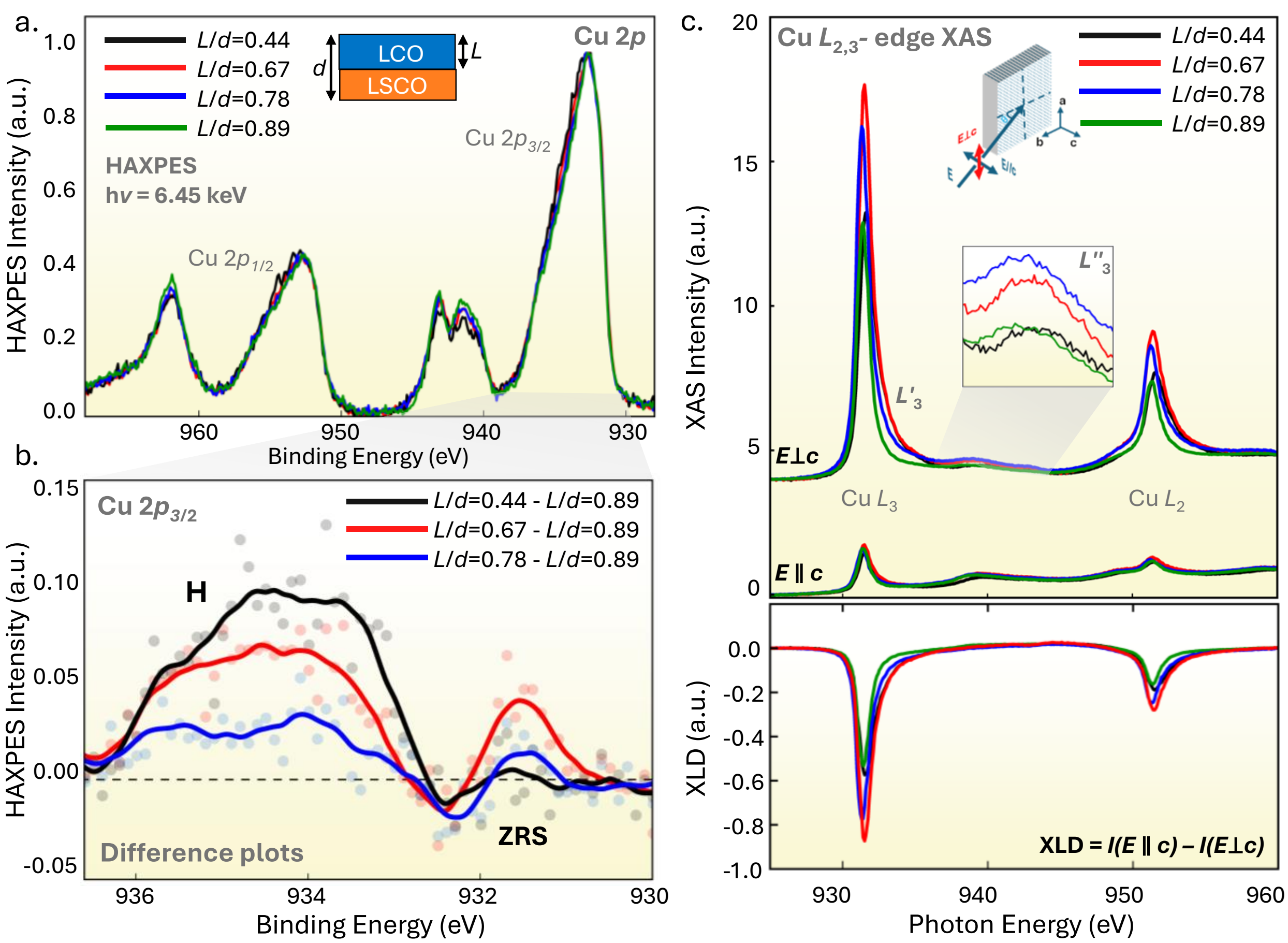

**FIG. 5. (a)** Cu 2$p$ core-level HAXPES spectra of the AHTS series, showing the Cu 2$p_{3/2}$ and Cu 2$p_{1/2}$ components together with the associated satellite and screened features. **(b)** Difference spectra of the Cu 2$p_{3/2}$ main-peak region, referenced to the $L/d$ = 0.89 spectrum, highlighting the doping-dependent evolution of Feature H and the ZRS-like screening shoulder. **(c)** Polarization-dependent Cu $L_{2,3}$-edge XAS spectra measured in luminescence-yield mode for the AHTS series. The upper panel shows spectra acquired with E⊥c and E∥c polarizations, while the lower panel shows the corresponding X-ray linear dichroism (XLD), highlighting enhanced unoccupied-state spectral weight and orbital polarization near the top of the superconducting dome. The inset enlarges the higher-energy Cu $L_3$-edge spectral-weight region.

Doping-induced spectral changes are highlighted by the Cu 2$p$ difference spectra [Fig. 5(b)] referenced to the $L/d$ = 0.89 sample with very low hole doping. This analysis reveals two distinct signatures of doping-dependent screening: a high-binding-energy satellite, Feature H, and a lower-binding-energy shoulder associated with $3d^9\underline{L}$, referred to as Zhang-Rice singlet (ZRS)-like screening [33]. Consistent with the superconducting $T_C$ dome, the ZRS shoulder is significantly enhanced at intermediate $L/d$ values but suppressed in both the underdoped ($L/d$ = 0.89) and overdoped ($L/d$ = 0.44) limits. The ZRS shoulder is associated with enhanced non-local screening through ligand-hole states, whereas Feature H reflects more poorly screened final states. Within the BPV framework, the observed nonmonotonic evolution is consistent with screening efficiency being governed by the proximity of the $\mu$ to the VHS [15,34], where the enhanced DOS identified in the previous section is expected to strengthen both local and non-local screening channels. At the underdoped and overdoped extremes, where the $\mu$ is expected to lie away from the VHS, non-local screening is reduced, resulting in a strong Feature H and a weak ZRS shoulder [35]. Near the top of the superconducting dome, where the $\mu$ is expected to approach the VHS-like condition, both local and non-local screening channels are enhanced, suppressing Feature H and amplifying the ZRS shoulder [15,35]. Therefore, Cu $2p$ core-level HAXPES indicates that spectral-weight redistribution is enhanced near the Lifshitz/VHS regime, where the associated VHS is expected to enhance the available screening channels. Together with the $\mu$ evolution and DOS enhancement discussed above, these results indicate that screening is strongest near the top of the superconducting dome, consistent with the central predictions of the BPV framework.

### E. Orbital Polarization and Enhanced Two-Dimensionality

To further examine the electronic-structure modifications associated with enhanced DOS and screening near the top of the superconducting dome, we turned to Cu $L_{2,3}$-edge XAS and XLD measurements [Fig. 5(c)]. The main absorption white line corresponds to transitions in the undoped Mott-insulating configuration from the initial state Cu$2p^6 3d^9$O$2p^6$ to the final state $Cu2p^5 3d^{10} O2p^6$. The spectra also show a high-energy satellite, labeled $L'_3$ in Fig. 5(c), associated with doped-hole states, arising from core-level transitions from the initial state $Cu2p^6 3d^9 O2p^5$ to the final state $Cu2p^5 3d^{10} O2p^5$, commonly referred to as the $3d^9\underline{L}$ or ZR satellite [26]. The intensity of the main Cu $L_3$ white line is strongest for the $L/d$ = 0.67 and 0.78 samples, indicating a modification of the unoccupied electronic states near the top of the superconducting dome.

To quantify the orbital anisotropy, we calculated the X-ray linear dichroism (XLD) signal from the polarization-dependent Cu $L_{2,3}$-edge XAS spectra shown in Fig. 5(c) as $\mathrm{XLD} = I(E \parallel c) - I(E \perp c)$. The resulting XLD spectra, shown in the bottom panel, provide a measure of the orbital anisotropy of the Cu $3d$ states.

Polarization-dependent XAS and the corresponding XLD analysis reveal that the in-plane Cu $d_{x^2-y^2}$ orbital weight reaches approximately 87% for $L/d$ = 0.67 and 77% for $L/d$ = 0.78. These values are substantially higher than those observed for $L/d$ = 0.44 and 0.89, indicating enhanced in-plane orbital polarization near the top of the superconducting dome. Dominant Cu $d_{x^2-y^2}$ orbital character is widely recognized as a key requirement for high-$T_c$ pairing [36,37]. The enhanced in-plane orbital polarization is not limited to states immediately above $E_F$ but persists up to several eV above the Fermi level [see inset of Fig. 5(c)], where the spectral weight remains strongest for the $L/d$ = 0.67 and 0.78 superlattices.

The enhanced in-plane orbital polarization observed near the top of the superconducting dome provides further evidence for geometry-driven electronic confinement in these AHTS. These results are consistent with resonant electronic confinement associated with a Fano-Feshbach resonance in the optimal range $0.67 \leq L/d \leq 0.78$ , as predicted for AHTS [16]. Within the BPV framework, variation of the $L/d$ ratio is expected to tune high-$T_c$ multigap pairing in the BEC-BCS crossover regime through a quantum resonance for pair exchange between the first subband, characterized by a high Fermi energy in the BCS regime, and the second subband, located near the top of the VHS and characterized by a lower Fermi energy. The different $T_c$ values observed for the $L/d$ = 0.67 and 0.78 samples are consistent with the asymmetric superconducting dome predicted in Ref. [16] for the Fano-Feshbach resonance, centered at a doping of 0.15 but with its maximum shifted toward lower doping (~0.1).

Together with the observed chemical-potential evolution, DOS enhancement, and enhanced screening response, the orbital-polarization results complete a unified picture in which the superconducting dome is governed by geometry-driven tuning of the electronic structure.

## IV. SUMMARY AND CONCLUSIONS

Using HAXPES and polarization-dependent XAS, we established a comprehensive spectroscopic picture of the electronic-structure evolution across this series of superconducting AHTS. The structural quality of the superlattices was established by high-resolution XRD, while their composition and interfacial hole doping were characterized through wide-range and shallow-core-level HAXPES measurements together with polarization-dependent O K-edge XAS. The evolution of the chemical potential was determined from La $3d$ core-level shifts, revealing occupation of the second subband for the $L/d$ = 0.67 and 0.78 superlattices, consistent with their position near the top of the superconducting dome. In this same regime, bulk-sensitive valence-band HAXPES measurements revealed enhanced spectral weight near the Fermi level, indicating an enhanced density of states available for pairing near a VHS-like condition. Cu $2p$ core-level spectroscopy further revealed enhanced local and non-local screening signatures, while polarization-dependent Cu $L_{2,3}$-edge XAS and XLD measurements demonstrated enhanced in-plane orbital polarization. Importantly, for these same samples, the near-$E_F$ density-of-states enhancement, screening response, and orbital polarization exhibit their strongest signatures.

Taken together, these independent spectroscopic observables support a unified picture in which the superconducting dome is governed by geometry-driven tuning of the electronic structure. The observed evolution of the chemical potential, density of states, screening response, and orbital character is consistent with the predictions of the BPV framework, in which confinement-induced electronic reconstruction drives the system toward a VHS associated with an electronic topological transition involving the second subband, providing a pathway for engineering superconductivity through geometric control of the electronic structure.

## ACKNOWLEDGMENTS

U.M.J., S.S., and A.X.G. acknowledge support from the US Air Force Office of Scientific Research (AFOSR) under award number FA9550-23-1-0476. A.X.G. also gratefully acknowledges the support from the Alexander von Humboldt Foundation. A.B. and G.C. thank the Superstripes - APS Association and the CNR project DCM.AD006.562 “Functional Disorder in Materials and Complex Systems” for supporting this work. The authors also acknowledge the use of an XRD facility supported by the Laboratory for Research on the Structure of Matter and the NSF through the University of Pennsylvania Materials Research Science and Engineering Center (MRSEC) DMR-2309043.

## REFERENCES

1. L. Manna, *Nano Lett.* **23**, 9673 (2023).

2. A. Brinkman, M. Huijben, M. van Zalk, J. Huijben, U. Zeitler, J. C. Maan, W. G. van der Wiel, G. Rijnders, D. H. A. Blank, and H. Hilgenkamp, *Nat. Mater.* **6**, 493 (2007).

3. A. Ohtomo and H. Y. Hwang, *Nature* **427**, 423 (2004).

4. A. Tsukazaki, A. Ohtomo, T. Kita, Y. Ohno, H. Ohno, and M. Kawasaki, *Science* **315**, 1388 (2007).

5. A. Bianconi, High $T_c$ superconductors made by metal heterostructures at the atomic limit, European Patent No. EP0733271 (1996).

6. N. Reyren, S. Thiel, A. D. Caviglia, L. Fitting Kourkoutis, G. Hammerl, C. Richter, C. W. Schneider, T. Kopp, A.-S. Rüetschi, D. Jaccard *et al.*, *Science* **317**, 1196 (2007).

7. A. Gozar, G. Logvenov, L. F. Kourkoutis, A. T. Bollinger, L. A. Giannuzzi, D. A. Muller, and I. Bozovic, *Nature* **455**, 782 (2008).

8. S. Smadici, J. C. T. Lee, S. Wang, P. Abbamonte, G. Logvenov, A. Gozar, C. D. Cavellin, and I. Bozovic, *Phys. Rev. Lett.* **102**, 107004 (2009).

9. A. Bianconi, D. Innocenti, A. Valletta, and A. Perali, *J. Phys.: Conf. Ser.* **529**, 012007 (2014).

10. A. Ikeda, Y. Krockenberger, Y. Taniyasu, and H. Yamamoto, *ACS Appl. Electron. Mater.* **4**, 2672 (2022).

11. D. Samal, N. Gauquelin, Y. Takamura, I. Lobato, E. Arenholz, S. Van Aert, M. Huijben, Z. Zhong, J. Verbeeck, G. Van Tendeloo *et al.*, *Phys. Rev. Mater.* **7**, 054803 (2023).

12. N. Bonmassar, G. Christiani, U. Salzberger, Y. Wang, G. Logvenov, Y. E. Suyolcu, and P. A. van Aken, *ACS Nano* **17**, 11521 (2023).

13. J. Y. Shen, C. Y. Shi, Z. M. Pan, L. L. Ju, M. D. Dong, G. F. Chen, Y. C. Zhang, J. K. Yuan, C. J. Wu, Y. W. Xie, and J. Wu, *Nat. Commun.* **14**, 7290 (2023).

14. G. Campi, A. Alimenti, G. Logvenov, G. A. Smith, F. F. Balakirev, S.-E. Lee, L. Balicas, E. Silva, G. A. Ummarino, G. Midei *et al.*, *Phys. Rev. Mater.* **9**, 074204 (2025).

15. A. Valletta, A. Bianconi, A. Perali, G. Logvenov, and G. Campi, *Phys. Rev. B* **110**, 184510 (2024).

16. G. Logvenov, N. Bonmassar, G. Christiani, G. Campi, A. Valletta, and A. Bianconi, *Condens. Matter* **8**, 78 (2023).

17. M. V. Mazziotti, A. Valletta, R. Raimondi, and A. Bianconi, *Phys. Rev. B* **103**, 024523 (2021).

18. M. V. Mazziotti, A. Bianconi, R. Raimondi, G. Campi, and A. Valletta, *J. Appl. Phys.* **132**, 193908 (2022).

19. A. Valletta, A. Bianconi, A. Perali, and N. L. Saini, *Z. Phys. B* **104**, 707 (1997).

20. A. Perali, A. Bianconi, A. Lanzara, and N. L. Saini, *Solid State Commun.* **100**, 181 (1996).

21. S. Krishnia, Y. Sassi, F. Ajejas, N. Sebe, N. Reyren, S. Collin, T. Denneulin, A. Kovács, R. E. Dunin-Borkowski, A. Fert *et al.*, *Nano Lett.* **23**, 6785 (2023).

22. W. Lin, L. Li, F. Doğan, C. Li, H. Rotella, X. Yu, B. Zhang, Y. Li, W. S. Lew, S. Wang *et al.*, *Nat. Commun.* **10**, 3052 (2019).

23. A. X. Gray, C. Papp, S. Ueda, B. Balke, Y. Yamashita, L. Plucinski, J. Minár, J. Braun, E. R. Ylvisaker, C. M. Schneider *et al.*, *Nat. Mater.* **10**, 759 (2011).

24. T.-L. Lee and D. A. Duncan, *Synchrotron Radiat. News* **31**, 16 (2018).

25. A. T. Young, E. Arenholz, J. Feng, H. Padmore, S. Marks, S. Schlueter, E. Hoyer, N. Kelez, and C. Steier, *Surf. Rev. Lett.* **9**, 549 (2002).

26. C. T. Chen, L. H. Tjeng, J. Kwo, H. L. Kao, P. Rudolf, F. Sette, and R. M. Fleming, *Phys. Rev. Lett.* **68**, 2543 (1992).

27. W. M. Li, J. F. Zhao, L. P. Cao, Z. Hu, Q. Z. Huang, X. C. Wang, Y. Liu, G. Q. Zhao, J. Zhang, Q. Q. Liu *et al.*, *Proc. Natl. Acad. Sci. U.S.A.* **116**, 12156 (2019).

28. M. Nagamoto, D. Ootsuki, T. Ishida, N. Tsutsumi, Y. Lee, A. Fujimori, A. Yasui, H. Kumigashira, S. Komiya, Y. Ando *et al.*, *Phys. Rev. B* **110**, 115150 (2024).

29. A. Ino, T. Mizokawa, A. Fujimori, K. Tamasaku, H. Eisaki, S. Uchida, T. Kimura, T. Sasagawa, and K. Kishio, *Phys. Rev. Lett.* **79**, 2101 (1997).

30. T. Tadano, Y. Nomura, and M. Imada, *Phys. Rev. B* **99**, 155148 (2019).

31. K. Okada and A. Kotani, *J. Phys. Soc. Jpn.* **58**, 2578 (1989).

32. A. Bianconi, M. De Santis, A. M. Flank, A. Fontaine, P. Lagarde, A. Marcelli, H. Katayama-Yoshida, and A. Kotani, *Physica C* **153-155**, 1760 (1988).

33. M. Taguchi, A. Chainani, K. Horiba, Y. Takata, M. Yabashi, K. Tamasaku, Y. Nishino, D. Miwa, T. Ishikawa, T. Takeuchi *et al.*, *Phys. Rev. Lett.* **95**, 177002 (2005).

34. R. P. Vasquez, D. L. Novikov, A. J. Freeman, and M. P. Siegal, *Phys. Rev. B* **55**, 14623 (1997).

35. M. A. Van Veenendaal, H. Eskes, and G. A. Sawatzky, *Phys. Rev. B* **47**, 11462 (1993).

36. H. Sakakibara, H. Usui, K. Kuroki, R. Arita, and H. Aoki, *Phys. Rev. Lett.* **105**, 057003 (2010).

37. M. Pompa, C. Li, A. Bianconi, A. Congiu Castellano, S. Della Longa, A. M. Flank, P. Lagarde, and D. Udron, *Physica C* **184**, 51 (1991).